\documentclass[11pt]{article}
\usepackage{moriond,epsfig}

\def\Journal#1#2#3#4{{#1} {\bf #2}, #3 (#4)}

\def\NIMA{{\em Nucl. Instrum. Methods} A}

\def\PLB{{\em Phys. Lett.}  B}
\def\PRL{\em Phys. Rev. Lett.}
\def\PRD{{\em Phys. Rev.} D}

\begin{document}
\vspace*{4cm}
\title{STUDIES OF TOP QUARK PROPERTIES
AT THE TEVATRON}

\author{VIATCHESLAV~SHARY \\
for the CDF and D0 collaborations}

\address{CEA, IRFU, SPP Centre de Saclay,\\
91191 Gif-sur-Yvette, France}

\maketitle\abstracts{
An overview of the recent measurements of the top quark properties
in proton antiproton collisions at $\sqrt{s}=1.96$ TeV is presented.
These measurements are based on  5.4 -- 8.7 fb$^{-1}$  of data collected with the D0 and CDF experiments at the Fermilab Tevatron collider. 
The top quark mass and width measurements, studies of the spin correlation in top quark pair production, W boson helicity measurement,
searches for anomalous top quark couplings and Lorentz invariance violation
are discussed.
}

\section{Introduction}

Top quark plays a special role in the modern particle physics.
Its large mass, close to 173~GeV, extremely short lifetime of about $5\cdot 10^{-25}$~sec,
large coupling to the Higgs boson and potentially large couplings to the beyond-standard-model particles
make the top quark an unique laboratory to test predictions of the standard model (SM)
and search for  phenomena beyond SM.
In this article the recent measurements of the top quark properties are reported, based on the 
5.4 -- 8.7 fb$^{-1}$  of data collected with the D0~\cite{d0:detector} and CDF~\cite{cdf:detector} experiments 
at the Fermilab Tevatron collider.

\section{Top Quark Mass Measurement}

Top quark mass is a free parameters in the SM. It is measured at the Tevatron collider mainly with a matrix element  or 
template based methods.
A significant reduction of the jet energy scale (JES) related uncertainties is achieved by using the W boson mass as a constrain 
for the invariant mass of two jets, produced by the W boson decay in the lepton+jet final state.
Recently the CDF collaboration measured the top quark mass using the full available statistics of 8.7~$fb^{-1}$ in
the lepton + jet final state~\cite{cdf:mass}.
In each events the reconstructed top mass  is found by minimizing a $\chi^2$ function
describing the overconstrained kinematics of the $t\bar{t}$ system for the combination with minimal $\chi^2$.
The improvement in precision could be reached by using 
a reconstructed top quark mass from 2nd best $\chi^2$ combination. 
Templates for the dijet mass of the hadronically decaying W boson are used to constrain the JES  in situ. 
The 3D probability density functions for different top quark masses
 are constructed using a kernel density estimation to the simulation. 
Comparison of the simulation templates with data ones allows to  extract the top quark mass.
This measurement results to the most precise determination of the top quark mass 
in a single channel: $ M_{top} = 172.85 \pm 0.71\ (stat.) \pm 0.84\ (syst.)~GeV$.
The precision of the measurement is limited mainly by the signal simulation uncertainties and residual JES uncertainty.

The D0 collaboration produced the new measurement in the dilepton channel using 5.4 $fb^{-1}$ of data using
the neutrino weighting technique to reconstruct the underconstrained kinematic in this channel~\cite{d0:mass}.
This measurement has slightly less statistical precision than the previous matrix element measurement~\cite{d0:me_mass},
but improved systematic uncertainty. The improvement has been achieved by propagating the in situ JES correction factor from
the lepton+jet top quark mass measurement~\cite{d0:ljet_mass} to the dilepton channel. 
Additionally the uncertainty on the b quark JES has been  reduced
by applying the specific JES corrections for the different type of partons in the simulation. These corrections 
reflect the difference in the signal particle response between data and MC for the gluon, light quark or b-quark jets.
The result of this measurement is
$M_{top} = 174.0 \pm 2.4\ (stat.) \pm 1.4\ (syst.)~GeV$.

The most recent combination of the top quark mass measurements at the Tevatron~\cite{comb:mass} 
doesn't include these two measurements yet, but has a remarkable precision of 
0.54\% and yield the top quark mass $M_{top} = 173.18 \pm 0.94~GeV$ (Fig.~\ref{fig:top_mass}).
The further improvement in the top quark mass uncertainty requires better understanding of the 
model uncertainties in the $t\bar{t}$ production simulation.

\section{Top Quark Width}
The top quark decay width ($\Gamma_t$) is a difficult parameter for the direct measurement. 
CDF collaboration made an attempt to measure the 
top quark width by comparing the reconstructed top mass distributions in simulation and data. Unfortunately the precision of this method 
allows only to establish the upper limit 
of about $\Gamma_t < 7.6$ GeV at  95\% C.L.~\cite{Aaltonen:2010ea} using luminosity of 4.3 $fb^{-1}$.

The D0 collaboration use the indirect determination  of the top quark decay width using 
the partial decay width $\Gamma (t \to Wb)$ and the branching fraction $B(t \to Wb)$.
The partial decay width is obtained from the t-channel single top quark production cross section as 
$\Gamma(t\to Wb) = \sigma(t-channel) \frac{\Gamma(t\to Wb)_{SM}}{\sigma(t-channel)_{SM}}$.
The total decay width can be written in terms of the partial decay width and the branching fraction 
as $\Gamma_t = \frac{\Gamma (t \to Wb)}{B(t \to Wb)}$.
Combining these two equations, the total decay width becomes
$\Gamma_t =  \frac{\sigma(t-channel) \Gamma(t\to Wb)_{SM}}{B(t \to Wb) \sigma(t-channel)_{SM}}$.
Using the $B(t \to Wb)$ measured value~\cite{d0:R} and t-channel cross section measurement~\cite{d0:single_top}, 
the resulting width is found to be $\Gamma_t =  2.00^{+0.47}_{-0.43}$~GeV for the top quark mass 172.5 GeV~\cite{Abazov:2012vd}.
This value could be reinterpret as a top quark lifetime $\tau_t = \hbar / \Gamma_t = 3.29^{+0.90}_{-0.63}\times 10^{-25}$~s. 

We can also use the measured value of $\Gamma_t$ to probe
the $Wtb$ interaction and directly determine the Cabibbo-Kobayashi-Maskawa  (CKM) 
quark mixing matrix element $|V_{tb}|$. Similar to the determination of $|V_{tb}|$ using the 
single top cross section, this determination  does not assume the unitarity of the CKM matrix 
or three generations of quarks. Additionally, using the measured branching fraction $B(t \to Wb)$ and 
t-channel as a single top cross section, allows to avoid an assumption that the top quark decays
exclusively to Wb and an assumption that the relative production
rate of s and t single top channels is the same  as predicted by the SM. 
Restricting the $|V_{tb}|$ values to the physically allowed region 
$0 \le |Vtb|^2 \le 1$, a lower limit on $|V_{tb}|$ matrix element could be establish:
$|Vtb| > 0.81$ at 95\% C.L.~\cite{Abazov:2012vd}.

\section{Spin Correlation in $t\bar{t}$ Production}

The unique feature of the top quark that it decays before depolarization or
hadronization. This peculiarity allows to study the spin related quantities via the 
top quark decay products, W boson and b quark.
Although top and antitop quarks are produced unpolarized
at hadron colliders, they have a significant correlation between
the orientation of theirs spins. The strength of spin correlation
depends on the production mechanism and on the calculation basis and 
could be measured via the angular distributions of top quark products.
The $t\bar{t}$ spin correlation strength C is defined by
$d\sigma_{t\bar{t}}^2/(d\cos\theta_1 d\cos\theta_2) = \sigma_{t\bar{t}} (1 - C \cos\theta_1 \cos\theta_2) / 4$, 
where $\sigma_{t\bar{t}}$ is a  production cross section, 
$\theta_1, \theta_2$ are  the angles between the spin-quantization axis and the
direction of flight of the down-type fermion from the W boson decay in the respective parent
$t$ or $\bar{t}$ rest frame.
Using the beam momentum vector as the quantization axis,
the SM predicts $C = 0.78 ^{+0.03}_{-0.04}$ at NLO QCD for the proton antiproton collision 
with the center-of-mass energy 1.96~TeV~\cite{spin:pred}.

Two approaches have been developed at the Tevatron for this measurement. The first one is the template based measurement,
which consist in the comparing the angular distributions templates with different spin correlation strength in simulation with data 
distribution. 
With this approach D0  measured in the dilepton channel
$C = 0.10 \pm 0.45 (stat.+syst.)$~\cite{d0:spin_temp}.
CDF measured in the dilepton channel $C = 0.04 \pm 0.56 (stat.+syst.)$~\cite{cdf:spin_dilep} and 
in the lepton+jet channel $C = 0.72 \pm 0.69 (stat.+syst.)$~\cite{cdf:spin_ljet}. All these results are compatible with SM prediction,
but have the limited statistical precision and can't distinguish between ``no spin correlation'' and ``spin correlation'' cases.

\begin{figure}
\begin{minipage}[tb]{.42\textwidth}
\includegraphics[width=\textwidth]{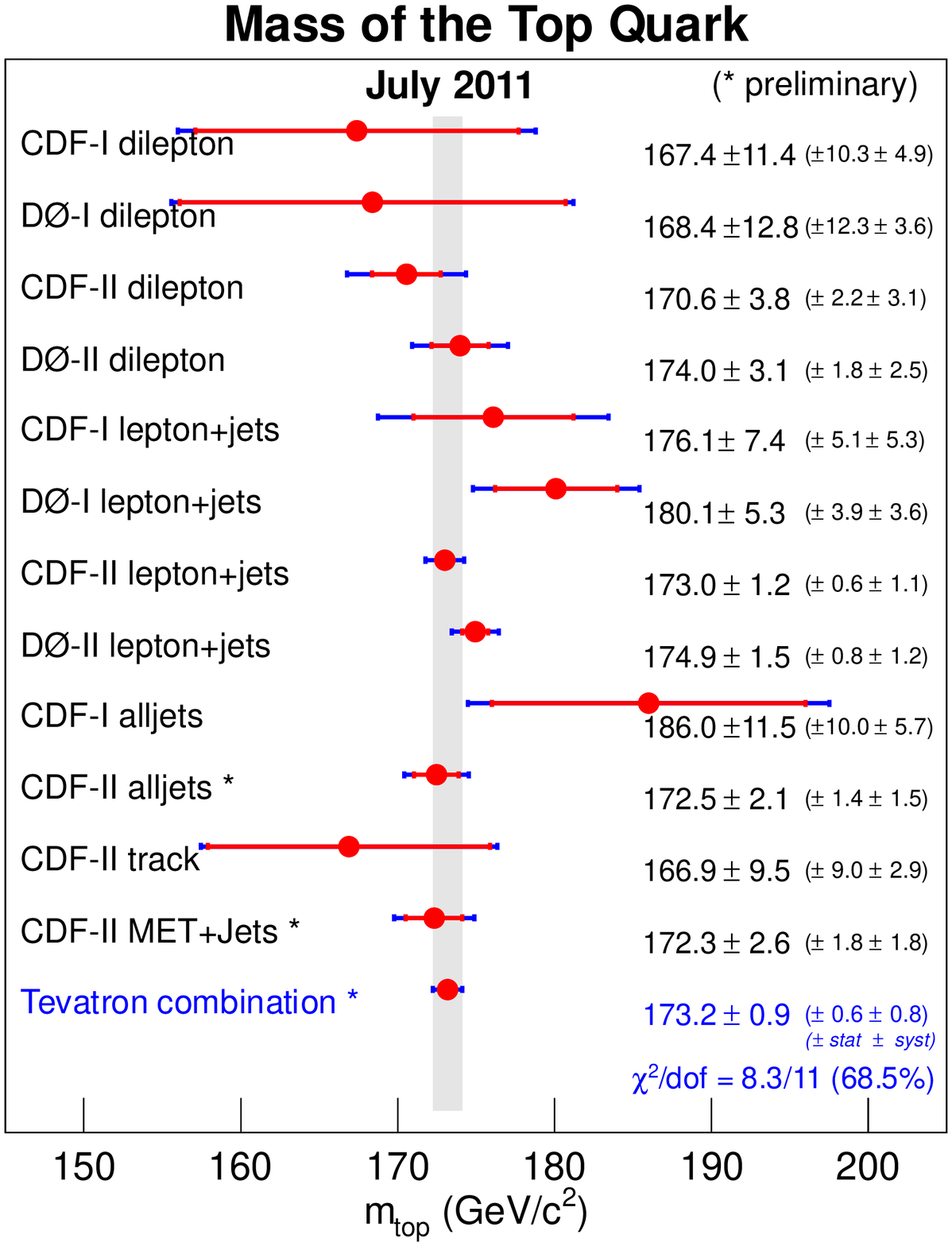}
\caption{Top quark mass combined result from D0 and CDF collaborations and results used in the combination.
\label{fig:top_mass} }
\end{minipage}
\hfill
\begin{minipage}[tb]{.56\textwidth}
\includegraphics[width=\textwidth]{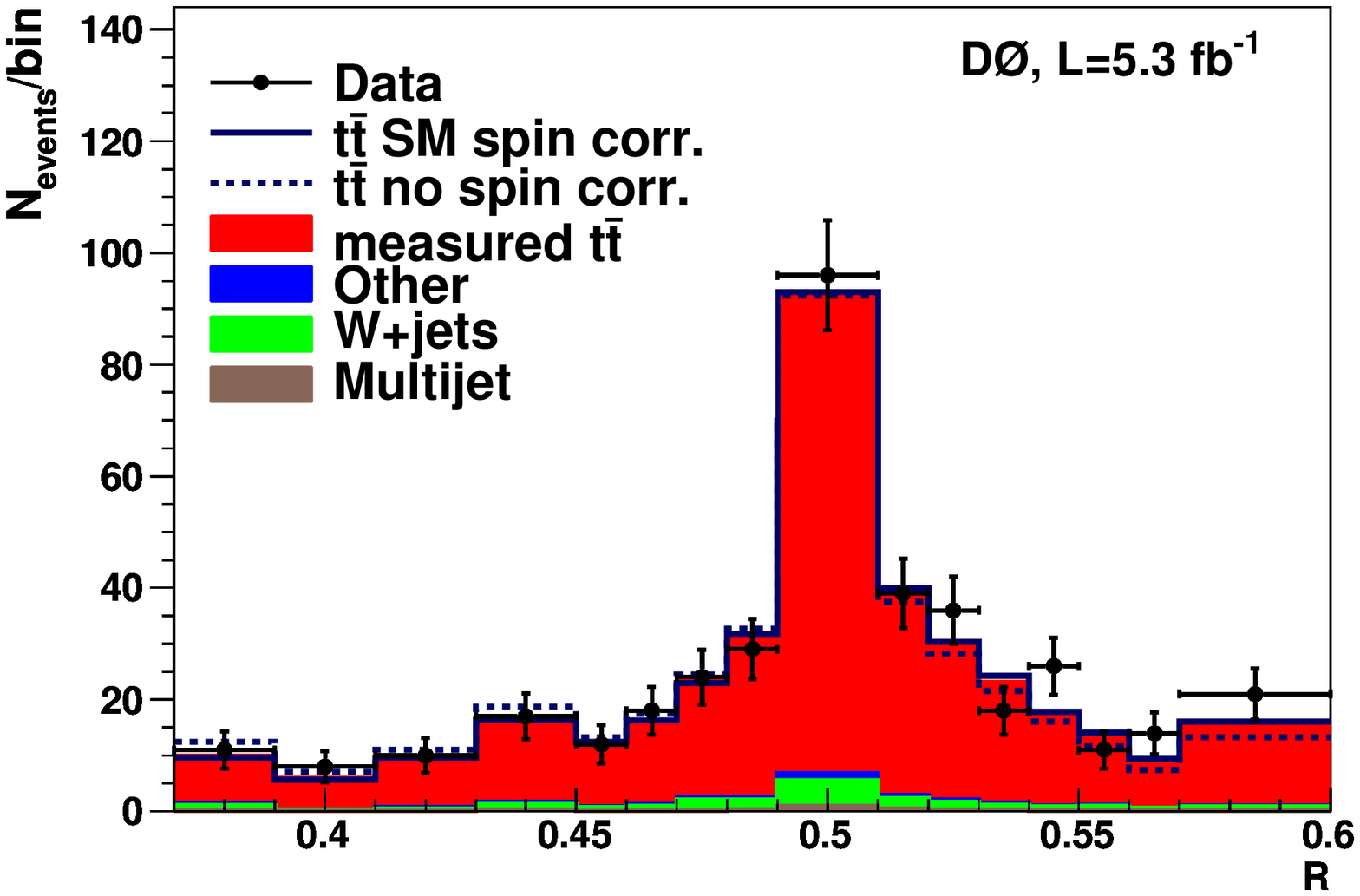}
\caption{The distribution of the discriminant R of the l+jets events with four jets.
The expectation (including background) for complete spin correlation
as predicted by the SM and the case of no spin correlation.
The measured $t\bar{t}$ distribution  is shown with the spin correlation strength measured in the 
dilepton and lepton+jet channels. 
\label{fig:spin_corr} }
\end{minipage}
\end{figure}

D0 pioneered the matrix element method for the spin correlation measurement. This method consist in the calculating the 
discriminant R for each event $R = \frac{P(x, H=1)}{P(x, H=c) + P(x, H=u)}$, where P(x) is a probability, 
calculated as a function of all measured kinematic variables $x$ using the leading order matrix element for the $t\bar{t}$ 
production for the ``spin correlation'' hypotheses ($H=c$) and ``no spin correlation'' hypothesis ($H=u$).
The simulated distributions of the discriminant R  for the correlation and no correlation hypotheses  are used 
to fit the distribution in data and determine the correlation strength~(Fig.~\ref{fig:spin_corr}).
Due to the maximal use of kinematic  information in the event and optimal use of the statistical weight of each event, this method
is 30\% more precise than the template based one. Measuring the spin correlation strength in the dilepton and lepton+jet channel and
combining them together, D0 obtained $C=0.66 \pm 0.23 (stat.+syst.)$ in agreement with the SM~\cite{d0:spin_me}.
This measurement has excluded the ``no spin correlation'' hypothesis for the first time a the level of $3.1\sigma$ 
or $C > 0.26$ at 95\% C.L.

\section {W Helicity Studies and Search for the Anomalous Quark Couplings}  

\begin{figure}
\centerline{\includegraphics[width=.7\textwidth]{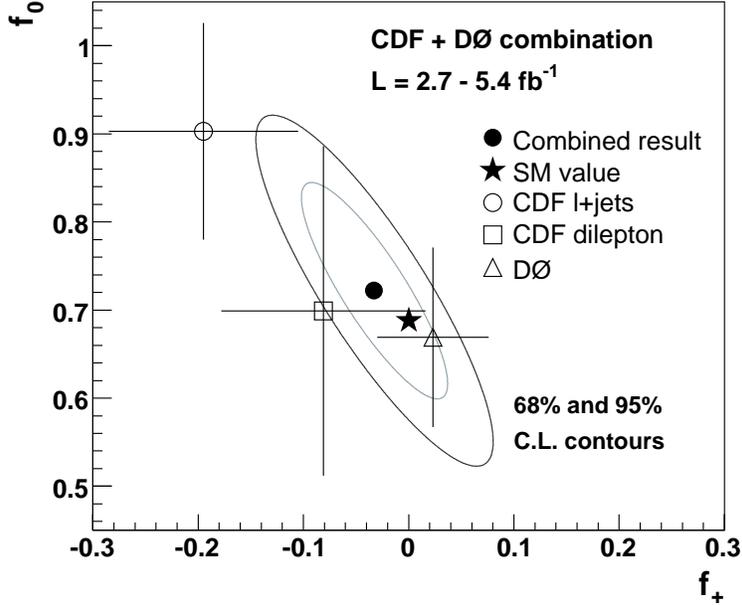}}
\caption{Combination of the 2D W boson helicity measurements. 
The ellipses indicate the 68\% and 95\% C.L. contours, 
the dot shows the best-fit value, and the star
marks the expectation from the SM. The input measurements
to the combination are represented by the open circle, square,
and triangle, with error bars indicating the 1 uncertainties
on $f_0$ and $f_+$. Each of the input measurements uses a central
value of mt = 172.5 GeV.
\label{fig:whel} }
\end{figure}

In the SM, the top quark decays, almost always, 
to a $W$ boson and $b$ quark via the $V-A$ charge current interaction. 
The new physics contribution may alter the fraction of the 
$W$ boson produced in different polarization states from SM values of 
$0.688\pm0.004$ for the longitudinal helicity fraction $f_0$,
$0.310\pm0.004$ for the negative helicity fraction $f_-$ and
$0.0017 \pm 0.0001$  for the positive helicity fraction $f_+$~\cite{Czarnecki:2010gb} at 
the top quark mass  of 173.3~GeV.
Both D0 and CDF collaborations have measured fractions $f_0$ and $f_+$
using the angular distribution of the down-type decay products of the $W$ boson 
(charged lepton or d, s quarks) in the rest frame of the $W$ boson.
Recently all these results with 2.7 -- 5.4 fb$^{-1}$ of integrated luminosity have been combined together~\cite{Aaltonen:2012rz}.
The obtained results are 
$f_0 = 0.722 \pm 0.081\ [0.062 (stat.) \pm  0.052 (syst.)]$
$f_+ = -0.033 \pm 0.046\ [0.034 (stat.) \pm 0.031 (syst.)]$
for measurements in which both $f_0$ and $f_+$ are varied
simultaneously (see Fig.~\ref{fig:whel}). These are the most precise measurements of $f_0$ and $f_+$ to date. 
The results are consistent with expectations
from the SM and provide no indication of new physics in
the $tWb$ coupling.

D0 also make a direct search for the anomalous coupling in $tWb$ vertex. 
Search for the phenomena beyond the SM  done in the form
of right-handed vector couplings $f_V^R$ or left- or right-handed
tensor couplings $f_T^L$, $f_T^R$, described by the effective Lagrangian
including operators up to dimension five:
\[
\mathcal{L} = - \frac{−g}{\sqrt{2}}\bar{b}\gamma^\mu V_{tb}(f_V^L P_L + f_V^R P_R)\ t W_\mu^- 
- \frac{−g}{\sqrt{2}}\bar{b}\ \frac{i\sigma^{\mu\nu} q_\nu V_{tb}}{M_W}
(f^L_T P_L + f^R_T P_R)\ t W^-_\mu + h.c.\ ,
\]
where $M_W$ is the mass of the $W$ boson, $q_\nu$ is its four-momentum, 
$V_{tb}$ is the Cabibbo-Kobayashi-Maskawa matrix
element, and $P_L = \frac{1}{2}(1 - \gamma_5)$ ( $P_R = \frac{1}{2}(1 + \gamma_5)$ ) 
is the left-handed (right-handed) projection operator. 
It is assumed that CP is conserved at the $Wtb$ vertex, 
meaning that all form factors $f^{L,R}_i$ are taken to be real. 
It is  also assumed that the top quark has spin $\frac{1}{2}$.
Variations in the coupling form factors would
mainly manifest themselves in two distinct ways: 
by changing the rate and kinematic
distributions of electroweak single top quark production, 
and by altering the fractions of $W$ bosons from top
quark decay produced in each of the three possible helicity states. 
D0 combines information from the measurement
of $W$ boson helicity fractions~\cite{d0:whel}
with information from measurement of single top quark production~\cite{d0:singletop}.
No indication of the beyond SM phenomena are found, so  upper limits on 
on anomalous $tWb$ couplings have been set, see Table~\ref{table:couplings}.
\begin{table}
\caption{\label{table:couplings}
Observed upper limits on anomalous $tWb$ couplings
at 95\% C.L. from W boson helicity assuming $f^L_V = 1$, from
the single top quark analysis, and from their combination, for
which no assumption on $f^L_V$ is made.}
\centerline{
\begin{tabular}{l|c|c|c}
& W helicity only & Single top only & Combination \\ \hline
$|f^R_V|^2$ & 0.62 & 0.89 & 0.30 \\ \hline
$|f^L_T|^2$ & 0.14 & 0.07 & 0.05 \\ \hline
$|f^R_T|^2$ & 0.18 & 0.18 & 0.12 \\ \hline
\end{tabular}
}
\end{table}

\section{Search for violation of Lorentz invariance}
\begin{figure}
\begin{minipage}[tb]{.48\textwidth}
\includegraphics[width=\textwidth]{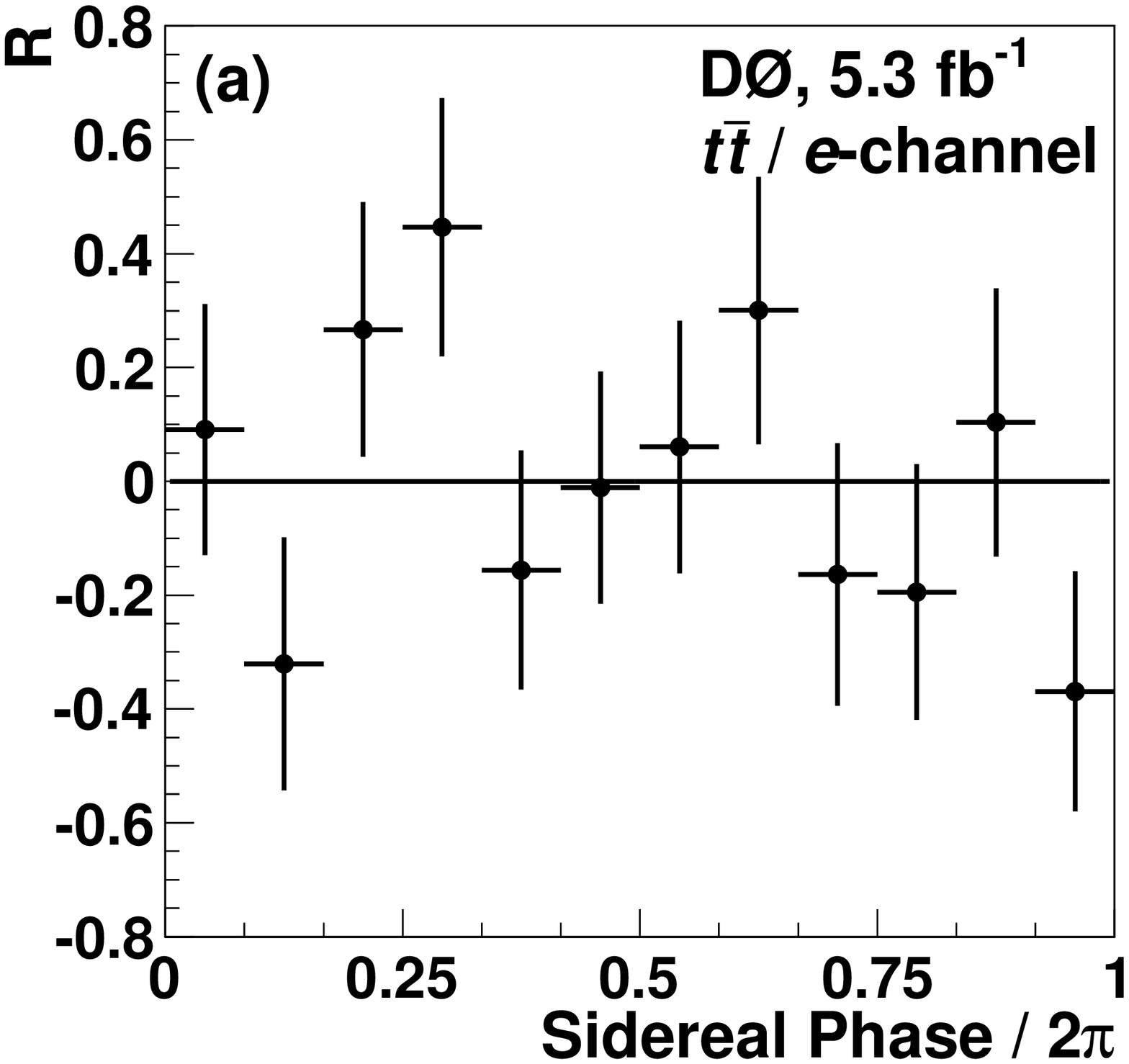}
\end{minipage}
\hfill
\begin{minipage}[tb]{.48\textwidth}
\includegraphics[width=\textwidth]{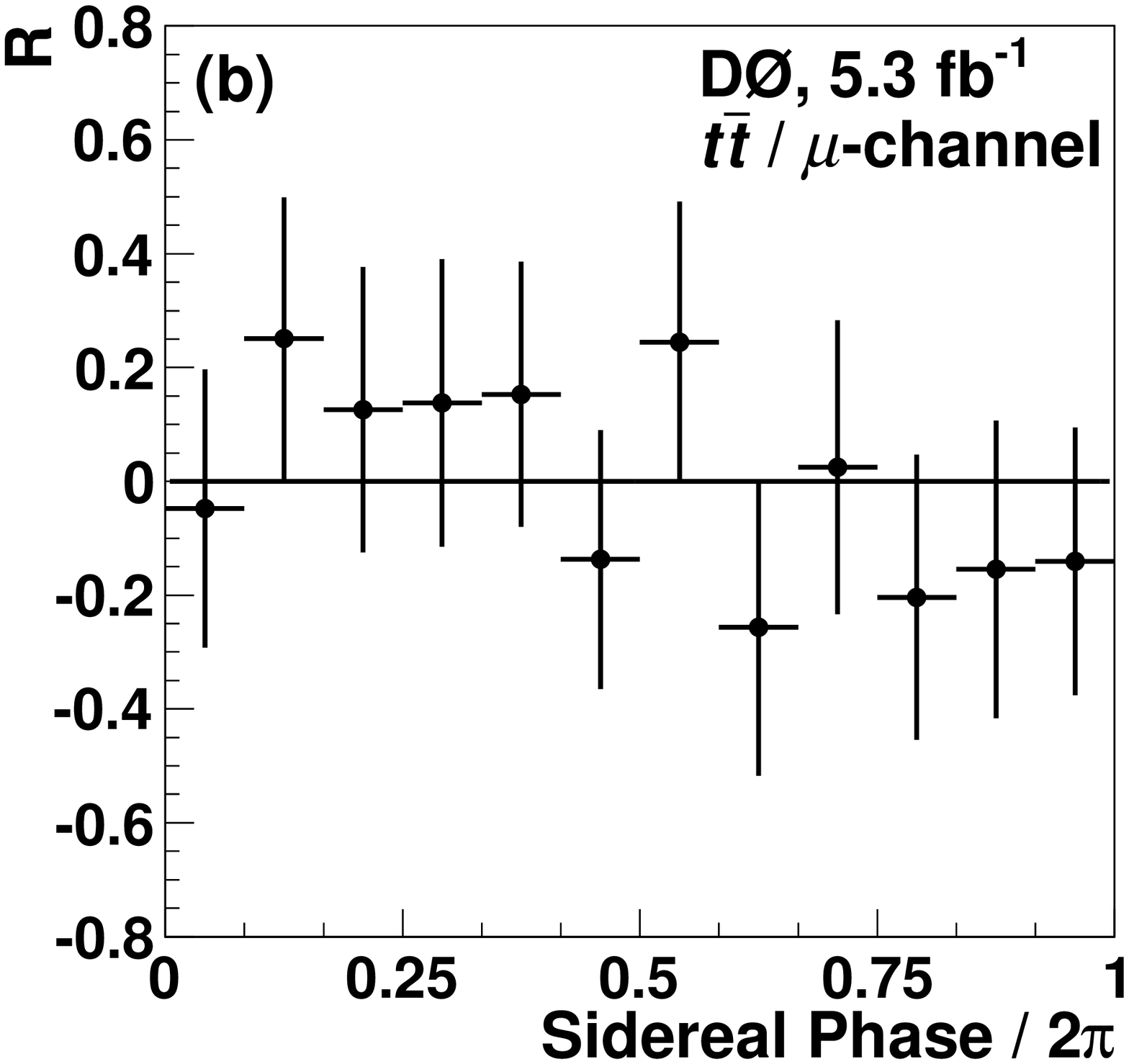}
\end{minipage}
\caption{The dependence of the normalized $t\bar{t}$ cross section R,
on the sidereal phase for e+jets candidates (left plot) and 
$\mu$+jets candidates (right plot).
\label{fig:liv}}
\end{figure}

D0 collaboration has searched for the violation of the Lorentz invariance in the $t\bar{t}$ production and decay
using 5.3$fb^{-1}$ of data in lepton+jet final state. 
The standard model extension  framework provides
an effective field theoretical treatment for violation of
Lorentz and CPT symmetry in particle interactions by
introducing Lorentz-violating terms to the Lagrangian
density of the SM~\cite{lv:theory}.
Rotation of the  Earth about its axis and about the Sun, leads to the time  dependence in the 
$t\bar{t}$ production cross section.
The relevant time scale is the sidereal day, 
which has a period of 23 hr 56 min 4.1 s.
Study of the $t\bar{t}$  cross section didn't reveal any time dependence with the sidereal period, see Fig.~\ref{fig:liv}.
In the absence of the any observed violation, the upper limit on the standard model extension has been set in 
the publication~\cite{d0:liv}.

\section{Conclusion}
The data collection at the Tevatron  has been stop at September 30, 2011. 
D0 and CDF collaborations accumulated about 10 $fb^{-1}$ of data.
Current results of the top quark properties measurement are based on the half of the available statistics mainly.
Both collaborations continue studies in the top quark sector, concentrating mainly on the measurement 
competitive with LHC (like a top quark mass measurement) or complementary to the LHC (like top quark production cross section 
and asymmetry measurement, spin correlation studies). More details about recent top quark results could be found on the 
collaborations web pages~\cite{d0:web,cdf:web}.


\section*{References}
\begin{thebibliography}{99}

\bibitem{d0:detector}
V.~M.~Abazov {\it et al.}, D0 Collaboration, \Journal{\NIMA}{565}{463}{2006}.

\bibitem{cdf:detector} 
  D.~Acosta {\it et al.}, CDF Collaboration,
  %``Measurement of the $t\bar{t}$ production cross section in $p\bar{p}$ collisions at $\sqrt{s} = 1.96$ TeV using lepton + jets events with secondary vertex $b-$tagging,''
\Journal{\PRD}{71}{052003}{2005}.

\bibitem{cdf:mass} 
CDF collaboration,  Conf Note 10761.

\bibitem{d0:mass}  
V.~M.~Abazov {\it et al.}, D0 Collaboration, Submitted to {\PRL},   arXiv:1201.5172.

\bibitem{d0:me_mass} 
V.~M.~Abazov {\it et al.}, D0 Collaboration, \Journal{\PRL}{107}{082004}{2011}.

\bibitem{d0:ljet_mass} 
V.~M.~Abazov {\it et al.}, D0 Collaboration, \Journal{\PRD}{84}{032004}{2011}.

\bibitem{comb:mass} 
  Tevatron Electroweak Working Group  for the CDF and D0 Collaborations,
  %``Combination of CDF and D0 results on the mass of the top quark using up to 5.8~fb-1 of data,''
  arXiv:1107.5255.

\bibitem{Aaltonen:2010ea}
  T.~Aaltonen {\it et al.}  CDF Collaboration,
  %``Direct Top-Quark Width Measurement CDF,''
  \Journal{\PRL}{105}{232003}{2010}.

\bibitem{d0:R}
V.~M.~Abazov {\it et al.}, D0 Collaboration, 
\Journal{\PRL}{107}{121802}{2011}.

\bibitem{d0:single_top}
V.~M.~Abazov {\it et al.}, D0 Collaboration,
\Journal{\PLB}{705}{313}{2011}.


\bibitem{Abazov:2012vd}
  V.~M.~Abazov {\it et al.}, D0 Collaboration,
  %``An Improved determination of the width of the top quark,''
  \Journal{\PRD}{85}{091104(R)}{2012}.

\bibitem{spin:pred}
W.~Bernreuther, A.~Brandenburg, Z.~G.~Si, and P.~Uwer,
\Journal{Nucl. Phys. B}{690}{81}{2004}.

\bibitem{d0:spin_temp}
V.~M.~Abazov {\it et al.}, D0 Collaboration,
\Journal{\PLB}{702}{16}{2011}.

\bibitem{cdf:spin_dilep}
CDF collaboration, ConfNote 10719 (2011).

\bibitem{cdf:spin_ljet}
CDF collaboration, ConfNote 10211 (2010).

\bibitem{d0:spin_me}
V.~M.~Abazov {\it et al.}, D0 Collaboration,
\Journal{\PRL}{108}{032004}{2012}.


\bibitem{Czarnecki:2010gb}
 A.~Czarnecki, J.~G.~Korner and J.~H.~Piclum,
 %``Helicity fractions of W bosons from top quark decays at NNLO in QCD,''
  \Journal{\PRD}{81}{111503}{2010}

\bibitem{Aaltonen:2012rz}
T.~Aaltonen {\it et al.}, CDF and D0 Collaborations,
  %``Combination of CDF and D0 measurements of the $W$ boson helicity in top quark decays,''
\Journal{\PRD}{85}{071106}{2012}.

\bibitem{d0:whel}
V.~M.~Abazov {\it et al.}, D0 Collaboration, 
\Journal{\PRD}{83}{032009}{2011}.

\bibitem{d0:singletop}
V.~M.~Abazov  {\it et al.}, D0 Collaboration, 
\Journal{\PLB}{708}{21}{2012}.

\bibitem{d0:couplings}
V.~M.~Abazov {\it et al.}, D0 Collaboration,
  %``Combination of searches for anomalous top quark couplings with 5.4 fb^-1 of ppbar collisions,''
Submitted to  {\PLB}, arXiv:1204.2332.

\bibitem{lv:theory}
D.~Colladay and V.A.~Kosteleck\'y, 
\Journal{\PRD}{58}{116002}{1998}; \\
V.A.~Kosteleck\'y
\Journal{\PRD}{69}{105009}{2004}.

\bibitem{d0:liv}
V.~M.~Abazov {\it et al.}, D0 Collaboration,
%``Search for violation of Lorentz invariance in top quark pair production and decay,''
Submitted to {\PRL}, arXiv:1203.6106.

\bibitem{d0:web}
D0 collaboration web page with top quark studies results: \\
http://www-d0.fnal.gov/Run2Physics/top/top\_public\_web\_pages/top\_public.html

\bibitem{cdf:web}
CDF collaboration web page with top quark studies results: \\
http://www-cdf.fnal.gov/physics/new/top/top.html

\end{thebibliography}
\end{document}